\documentclass[
aps,prl,floatfix, twocolumn, 
superscriptaddress]{revtex4}
\usepackage{amsmath,amssymb}
\usepackage{graphicx}
\usepackage{epsfig}
\usepackage{ulem}
\usepackage[usenames]{color}

\begin{document}

\title{Electron-hole scattering-induced temperature behaviour of HgTe-based \\ semimetal quantum well }

\author{A.~V.~Snegirev}
\affiliation{Rzhanov Institute of Semiconductor Physics, \\
Siberian Branch of Russian Academy of Sciences, Novosibirsk 630090, Russia}
\affiliation{Novosibirsk State University, Novosibirsk 630090, Russia}

\author{V.~M.~Kovalev}
\affiliation{Rzhanov Institute of Semiconductor Physics, \\
Siberian Branch of Russian Academy of Sciences, Novosibirsk 630090, Russia}
\affiliation{Novosibirsk State Technical University, Novosibirsk 630073, Russia}

\author{M.~V.~Entin}
\affiliation{Rzhanov Institute of Semiconductor Physics, \\
Siberian Branch of Russian Academy of Sciences, Novosibirsk 630090, Russia}

\author{E.~B.~Olshanetsky}
\affiliation{Rzhanov Institute of Semiconductor Physics, \\
Siberian Branch of Russian Academy of Sciences, Novosibirsk 630090, Russia}

\author{N.~N.~Mikhailov}
\affiliation{Rzhanov Institute of Semiconductor Physics, \\
Siberian Branch of Russian Academy of Sciences, Novosibirsk 630090, Russia}

\author{Z.~D.~Kvon}
\affiliation{Rzhanov Institute of Semiconductor Physics, \\
Siberian Branch of Russian Academy of Sciences, Novosibirsk 630090, Russia}
\affiliation{Novosibirsk State University, Novosibirsk 630090, Russia}

\begin{abstract}
The semimetal quantum well (QW) based on HgTe structures exhibiting unusual
transport properties at low temperature is examined experimentally. It demonstrates either
a linear or quadratic growth of resistance with temperature at different top-gate
voltages in the semimetal regime. We develop a theoretical model of HgTe-based semimetal QW resistance temperature dependence based on electron-hole scattering processes at low temperatures.  We apply the Boltzmann transport equation approach to study the effect of electron-hole scattering in a semimetal QW. The calculated temperature behavior of 2D semimetal resistivity demonstrates an excellent agreement with experimental findings.
\end{abstract}

\maketitle
\textit{Introduction.---} Analysis of different scattering mechanisms under different temperature
regimes is a key ingredient for understanding the material conductivity temperature
dependence. At low temperature, the dominating scattering mechanisms are electron-impurity
and interparticle scattering. In Galilean invariant systems with a parabolic
electron dispersion, the electron-electron scattering does not affect the conductivity
due to the conservation of the total momentum of colliding particles. It becomes interesting to
study the systems where this law is violated, and the interparticle scattering becomes
pronounced and effectively influences the conductivity, at least in a low temperature
domain. Recently, quantum wells based on the CdHgTe/HgTe/CdHgTe structure
have been actively studied both experimentally and theoretically, mainly due to their
unique properties as toplogical insulators \cite{TI_1,TI_2,TI_3,TI_4,TI_5,TI_6}. The physical properties of
CdHgTe/HgTe/CdHgTe quantum wells (QW) are strongly determined by the QW width and orientation with respect
to the growth axis. It is experimentally proven that there is a critical QW width below
which the HgTe QW behaves as a topological insulator. In contrast, in thick enough HgTe QWs, valence
bands overlapped with a conduction band minimum, thus, resulting in the existence of
electrons and holes in the same QW, the semimetal regime 
\cite{100_c1,sm1,100_c2,100_c3,sm2}. Under the action of external uniform
electric field, holes and electrons move in the opposite directions and their collisions strongly affect
the HgTe QW conductivity at low temperatures. Here we experimentally demonstrate that, at low
temperatures, $T$, HgTe QW being in the semimetal regime, demonstrates either $T^2$ or linear in the $T$ behavior of the QW resistance depending on the top-gate voltage applied to the QW. We assume that these dependencies are from the effective electron-hole scattering in a sample. Below, we present the experimental results of HgTe QW resistivity temperature behavior measurements and develop a theoretical model that reasonably well explains these experimental data. 

\begin{figure}[t!]
	\centering{\includegraphics*[width=0.85\linewidth]
		{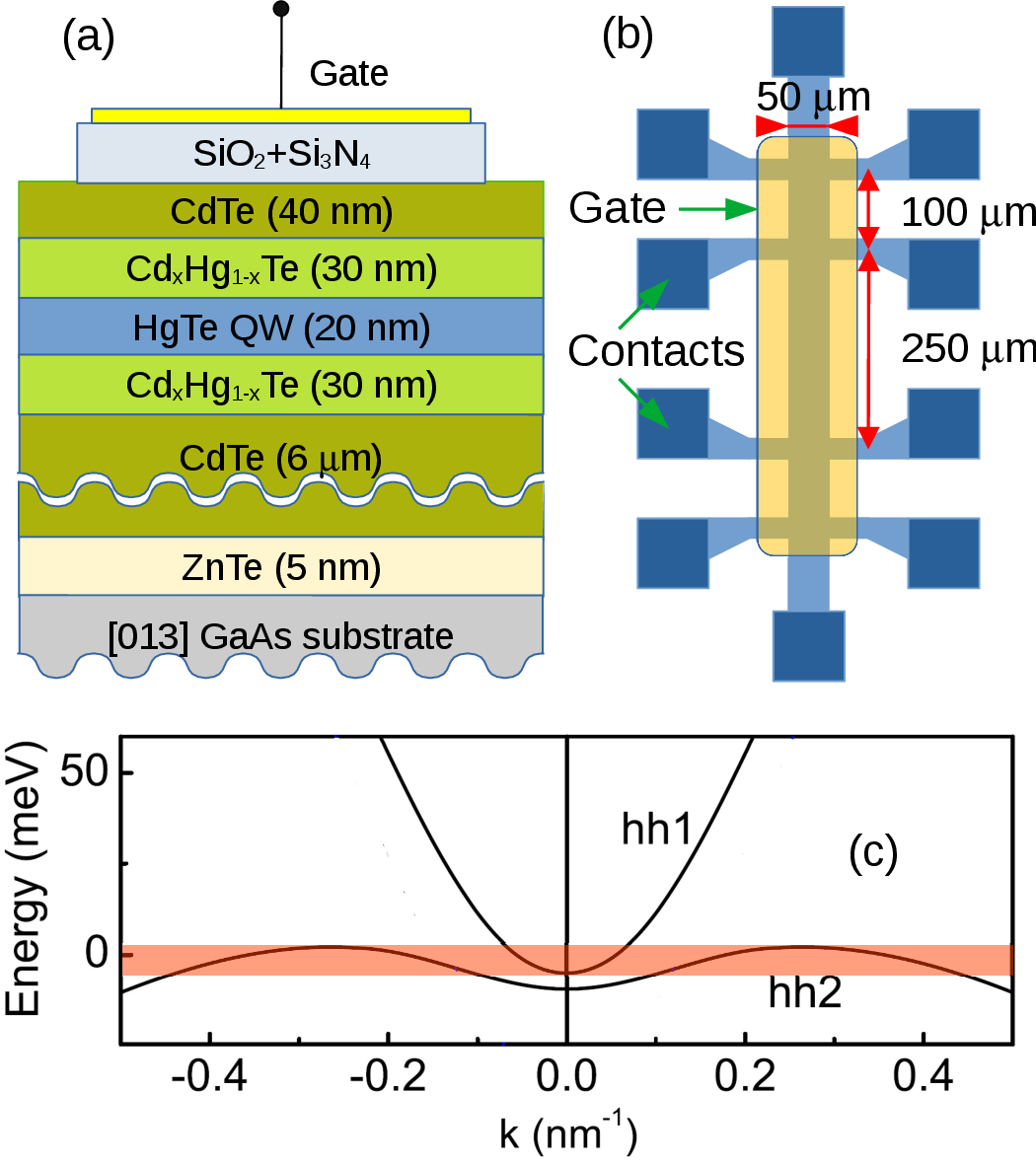}}
	\caption{
  {\bf (a)} Layer-by-layer structure of the quantum well; {\bf (b)} top view of the sample; {\bf (c)} schematic representation of the band spectrum of a 20 nm HgTe QW. The area where the conduction and valence bands overlap is highlighted.
}
\label{f:1}
\end{figure}

\textit{Experiment.---}The samples used in the experiment were made on the basis of 20 nm HgTe QWs with the (100) orientation, grown using the technology described earlier (\cite{100_c1,100_c2} and the references therein). A schematic section of the QW is shown in Fig.~\ref{f:1}a. The experimental samples were standard Hall bars equipped with an electrostatic gate with a width of $W=50\mu$m and a distance between voltage probes of $L=100\mu$m and $L=250\mu$m, Fig.~\ref{f:1}b. Using an electrostatic gate, the concentration of holes and electrons in the quantum well could be varied over a wide range. The measurements were carried out at temperatures of $(0.2-5)K$ and in magnetic fields up to 5 T using standard phase detection techniques. The signal frequency was 6-12 Hz, and the driving currents were at the order of 1-10 nA to avoid heating effects.

\begin{figure}[t!]
	\centering{\includegraphics*[width=0.85\linewidth]
		{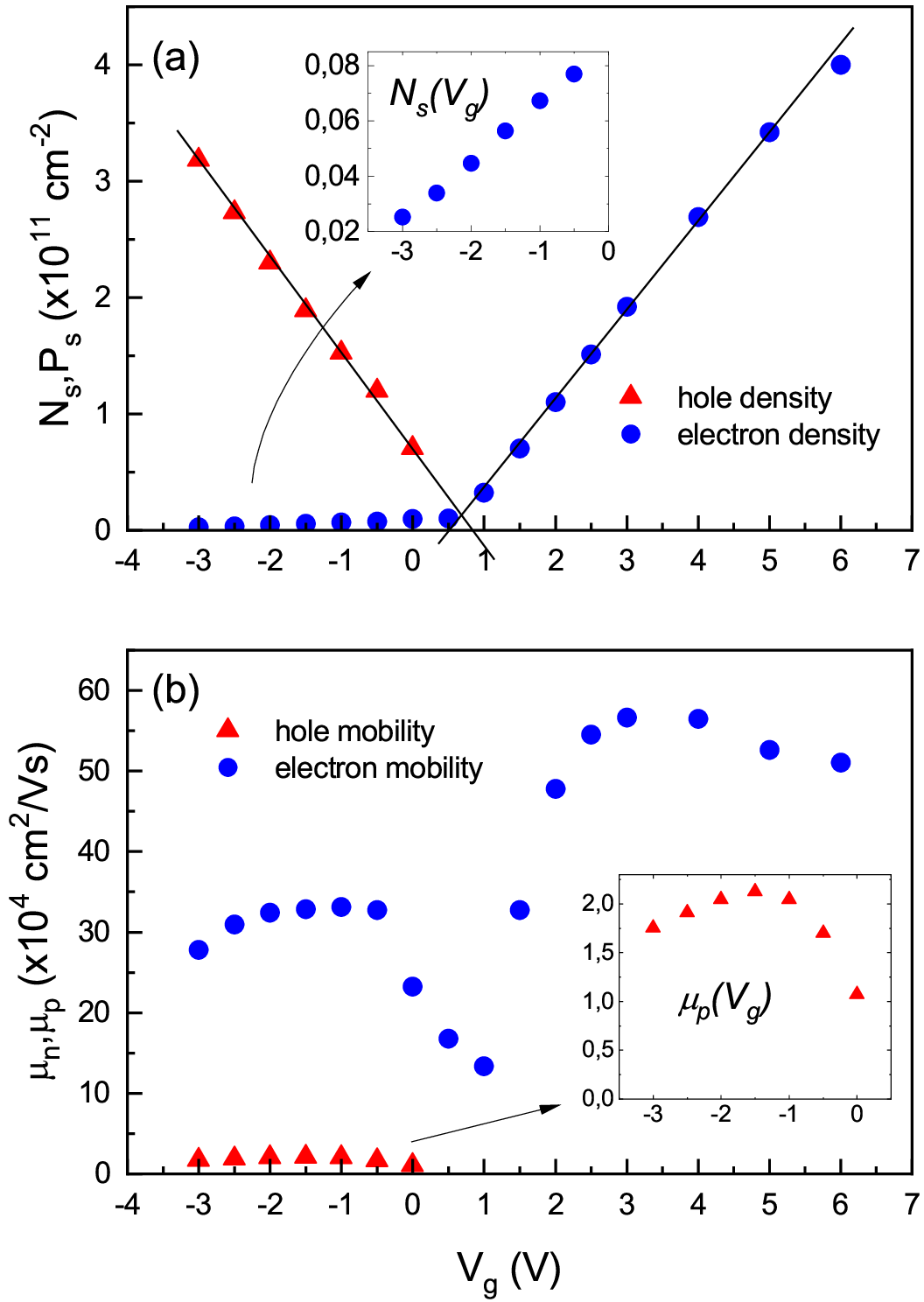}}
	\caption{
  (a) Gate voltage dependence of the
		density of 2D holes and 2D electrons in a 20nm (100) HgTe QW.
		Insert: magnified gate voltage dependence of the electron density
		in the interval from -0.5 to -3 V; (b) gate voltage dependence of
		the mobility of 2D holes and 2D electrons in the 20nm (100) HgTe
		QW. Insert: magnified gate voltage dependence of the hole mobility in
		the interval from 0 to -3 V.}
	\label{f:2}
\end{figure}

In Fig.~\ref{f:2} are the dependencies of the concentration and mobility of electrons and holes on the gate voltage obtained from the analysis of the behavior of $\rho_{xx}$ and $\rho_{xy}$ in weak magnetic fields. As can be seen, in the gate voltage range $-3 V\leq V_g \leq +0.7 V$ the QW contains both two-dimensional electrons and holes, i.e. a two-dimensional semimetal is realized. As in other previously studied wide HgTe QWs with surface orientations (013) and (112), the main reason for the formation of a semimetal in the 20 nm HgTe (100) QW is the indirect overlap of the conduction band and valence band, caused by the deformation of HgTe due to a small differences in lattice constants of HgTe and CdTe, Fig.~\ref{f:1}c. In a 20 nm HgTe(100) QW, this band overlap is relatively small, $1-1.5 $ meV. When the gate voltage changes from $+0.7 V$ to $-3 V$, the measured hole density to electron density ratio $P_s / N_s$ changes from 1 to approximately $10^{2}$. The concentrations of electrons and holes at the charge neutrality point (CNP) ($V_g=+0.7V$): $N_s = P_s\approx1.2 \times10^{10} cm^{-2}$, \cite{100_c3}.

\begin{figure}[t!]
	\centering{\includegraphics*[width=0.85\linewidth]
		{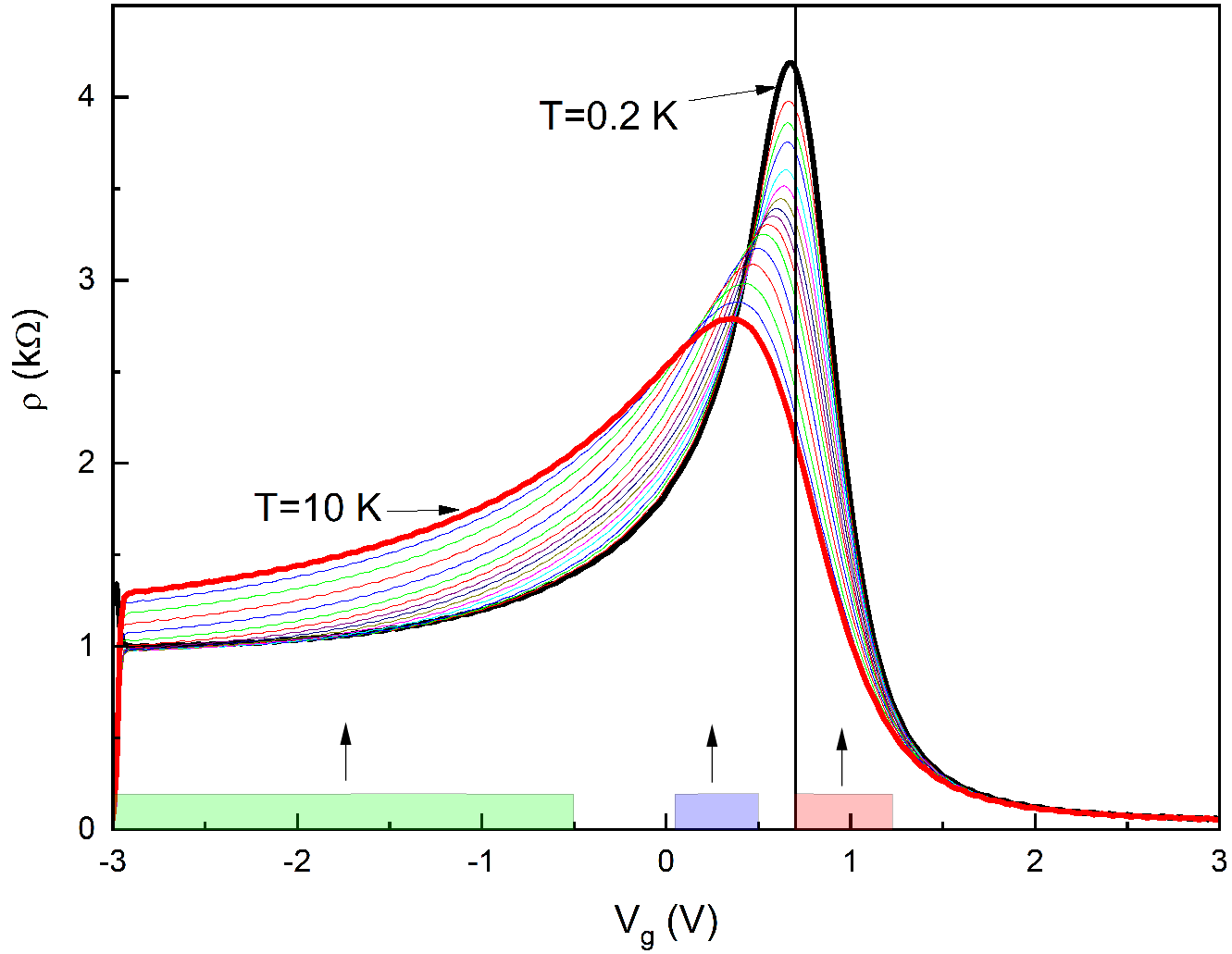}}
	\caption{
  Gate voltage dependence of resistivity $\rho$ at B=0 and temperatures T=0.2, 0.5, 0.7, 1, 1.5, 2, 2.5, 3, 3.5, 4.2, 5, 6, 7, 8, 9 and 10 K. The vertical line at $V_g=0.7 V$ marks the approximate position of the CNP.
	}
	\label{f:3}
\end{figure}
\begin{figure}[t!]
	\centering{\includegraphics*[width=0.85\linewidth]
		{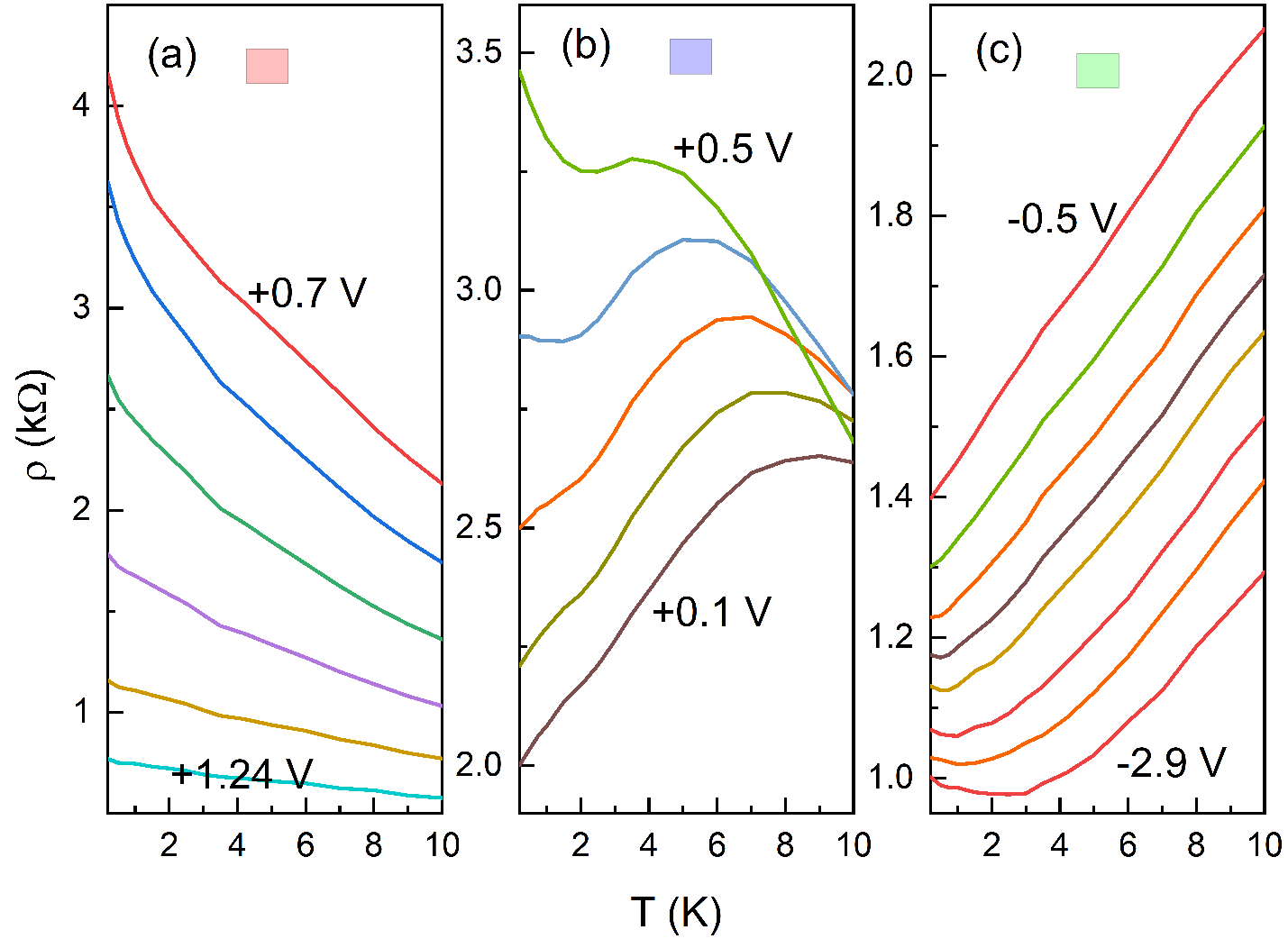}}
	\caption{
 Temperature dependence of resistivity $\rho$ at B=0 at different gate voltage ranges:
		(a) from 2DEG to CNP: $V_g$= 1.24, 1.1, 1.0, 0.9, 0.8 and 0.7 V.
		(b) region of T-dependence inversion: $V_g$= 0.5, 0.4, 0.3, 0.2, 0.1 V
		(c) semimetal, from linear to parabolic T-dependence: $V_g$= - 0.5, -0.7, -0.9, -1.1, -1.3, -1.7, -2.1 and -2.9 V.
	}
	\label{f:4}
\end{figure}

The resistivity dependence on the gate voltage at $B=0$ and in the temperature range from 0.2 K to 10 K is shown in Fig.~\ref{f:3}. In the $\rho(V_g)$ dependence,  there is a peak which, at low temperatures, coincides with the CNP, but with the increasing temperature it decreases in height and shifts to the left. In addition, as can be seen in Fig.~\ref{f:4}, when the gate voltage changes, an inversion of the temperature dependence of the resistance is observed. Thus, in the gate voltages region to the right of CNP, at which only a two-dimensional electron gas (2DEG) is present in the QW, the resistance increases with the decreasing temperature (Fig.~\ref{f:4}a), and the stronger this behavior the lower the electron concentration is. Then, in the gate voltages narrow region to the left of the CNP, an inversion of the temperature dependence is observed. The temperature dependence of resistance in this region is nonmonotonic (Fig.~\ref{f:4}b). Finally, at gate voltages from $-0.5\,V$ to $-3\,V$, in the region of existence of a semimetal with a predominance of holes over electrons, the dependence of resistance on temperature has the form characteristic of metallic conductivity, Fig.~\ref{f:4}a, i.e. the resistance is increased with the increasing temperature. At the same time, in the studied temperature range, at a certain interval of gate voltages, an almost linear temperature dependence of $\rho\sim T$ is first observed, which, at higher negative biases on the gate, is replaced by a quasiparabolic one. In the following part of the article we focus on the analysis of the $\rho(T)$ dependence precisely in this region of gate voltages corresponding to a two-dimensional semimetal.

\begin{figure}[b!]
\centering \includegraphics[width=0.85\columnwidth]{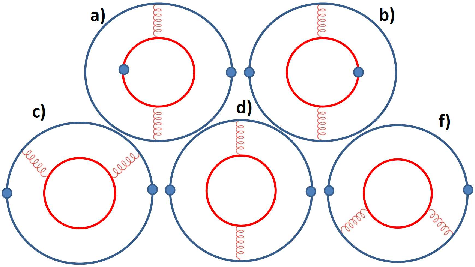}
\caption{Feynman diagrams describing the lowest order corrections to the
electron conductivity due to the electron-hole Coulomb scattering. Blue (red) lines are
the electron (hole) Green's functions, wavy lines are the Coulomb interaction.
Small blue circles are velocity vertices. Figures a) and b)
describe the electron-hole drag effect, c) and f) correspond to the density of states corrections, d) describes the distribution function correction.}
\label{plot2}
\end{figure}

\textit{Theoretical model.---} The theoretical model under study is a 2D electron-hole gas. We assume that the residual conductivity at zero temperature is determined by the particle-impurity
scattering, whereas the low-temperature corrections to the semimetal conductivity come mainly from highly-mobile electrons scattered off the low-mobile holes, as it is seen in Fig.\ref{f:2} in the gate voltages region $(-3V)\div(-0.5V)$, corresponding to the semimetal regime. It is also well known that
the electron effective mass is an the order of magnitude lower in comparison with
hole effective masses (see band structure of electrons and holes in
Fig.~\ref{f:1}c). At the same time, the hole density is much larger then the electron one, and is increased further with lowering top-gate voltages towards $-3V$, Fig.~\ref{f:2}a. 
Thus, it is possible to assume that the hole screening of
hole-impurities interaction is predominant -- because the polarization operator,
determining the screening length, is directly proportional to the density of
states, say, particle effective mass in 2D system -- and extremely sensitive to the top-gate voltage. 

Thus, the key idea of our theoretical model is as follows. At gate voltages $\sim(-0.5V)$ the hole density is low enough, but much larger then an electron one, (Fig.~\ref{f:2}a), and holes move diffusively, $T\tau_{h}<1$ (here and below, $\tau_{h,e}$ are hole and electron momentum relaxation times due to the hole-impurity and electron-impurity scattering, respectively), and electrons scattered off these holes demonstrate the linear-in-$T$ correction to the sample resistivity (Fig.~\ref{f:4}c). With a further gate voltage lowering till $\sim(-3V)$, the hole density is increased and holes effectively screen of hole-impurity scattering, resulting in the drop of zero temperature resistivity and converting the hole motion to the quasiballistic regime, $T\tau_{h}>1$. Thus, electrons scattered off the quasiballistic holes result in the quadratic-in-$T$ temperature correction to the sample resistivity at gate voltage $\sim(-3V)$, (Fig.~\ref{f:4}c). At the same time, we assume that
electrons, being highly-mobile particles (Fig.~\ref{f:2}b) are in the quasiballistic regime ($T\tau_{e}\gg1$) of their motion over all the range of gate voltages corresponding to the semimetal regime, i.e., from $-0.5\,V$ to $-3\,V$.

At low temperatures, when the relaxation rate of particle momentum  due to the scattering on impurities, is $1/\tau_{e,h}\gg g T^2/E_F$, (here, $g T^2/E_F$ is a particle-particle collision rate, with $g$ being the effective particle-particle interaction constant), the interparticle Coulomb scattering can be considered as a correction to the
residual conductivity determined by the particle collisions with impurities. In this
Letter we consider the lowest order corrections on the electron-hole Coulomb
interaction potential to the electron conductivity and resistivity. The
corresponding Feynman diagrams describing the lowest order corrections to the
electron conductivity due to the electron-hole Coulomb scattering are given in
Fig.~\ref{plot2}. Figs. \ref{plot2}a and \ref{plot2}b describe the electron-hole drag effect, \ref{plot2}c and \ref{plot2}f
are of the density of states corrections, whereas \ref{plot2}d describes the distribution function
correction. A proper account of all these diagrams can be made via the Boltzmann transport
equation, as it is shown in \cite{Maslov}. Below, we apply this approach to
calculate the low temperature corrections to the electron conductivity induced
by the electron scattering off holes.

\textit{Theory of electron conductivity correction caused by Coulomb
interaction with holes.---}We start from the Boltzmann equation for electron distribution function, $f_{\bf
p}$, written as
\begin{gather}\label{q1}
\frac{\partial f_{\bf p}}{\partial t}+{\bf F}\cdot\frac{\partial f_{\bf
p}}{\partial {\bf p}}
=Q_{i}\{f_{\bf p}\}+Q_{eh}\{f_{\bf p}\},
\end{gather}
where ${\bf F}=e{\bf E}$ is a static force affecting to electrons. The electron-impurity collision integral is $Q_{i}\{f_{\bf p}\}=2\pi n_iu_0^2\sum_{{\bf p}'}(f_{{\bf p}'}-f_{\bf p})
\delta(\epsilon_{{\bf p}'}-\epsilon_{\bf p})$, whereas the electron-hole collision integrals read
\begin{gather}\label{q2}
Q_{eh}\{f_{\bf p}\}=2\pi\sum_{{\bf p}',{\bf k}',{\bf k}}|U_{{\bf p}'-{\bf p}}|^2
\Bigl[(1-f_{\bf k})(1-f_{\bf p})f_{{\bf k}'}f_{{\bf p}'}-
\\\nonumber
(1-f_{{\bf k}'})(1-f_{{\bf p}'})f_{{\bf k}}f_{{\bf
p}}\Bigr]
\delta(\epsilon_{{\bf k}'}+\epsilon_{{\bf p}'}-\epsilon_{{\bf k}}-\epsilon_{{\bf
p}})
\delta_{{\bf k}'+{\bf p}'-{\bf k}-{\bf p}}.
\end{gather}
Here ${\bf p}$ refers to an electron momentum, and ${\bf k}$ is a hole momentum,
$n_i$ is an impurities density, $u_0$ is a strength of short-range impurity
potential, $u({\bf r})=u_0\delta({\bf r})$, $U_{\bf q}$ is Coulomb interaction
potential, $\epsilon_{\bf p}=p^2/2m_e$ is an electron dispersion and $\epsilon_{\bf
k}=({\bf k}\pm{\bf k}_0)^2/2m_h-\Delta$ is a hole dispersion, Fig.~\ref{f:1}c. 
We will ignore here the intervalley scattering both in the
particle-impurity and particle-particle scattering channels, thus, the shift of
hole valleys by $\pm{\bf k}_0$ does not play the role, and further we count the
holes momenta from this value, ${\bf k}\pm{\bf k}_0\rightarrow{\bf k}$. The
doubling of single-valley hole density due to two valleys will be accounted in the
total hole density.
Here and below, we put $\hbar=1$ and restore it in the final expressions. Further,
we consider the electron and hole scattering off short-range impurities, and the
corresponding scattering times are assumed to be independent from the
corresponding particle energy. In this case, we use the relaxation time
approximation for the electron-impurity and hole-impurity scattering processes $Q_{i}\{f_{\bf p(k)}\}=-(f_{\bf p(k)}-n_{\bf p(k)})/\tau_{e(h)}$. 
%
%
%
%
%
%
Here $n_{\bf p(k)}$ is an equilibrium Fermi distribution function of electrons
(holes), and $\tau_{e(h)}$ are particle-impurity momentum relaxation times. The
distribution function is expanded into the powers of external EM field as
$f_{\bf p}=n_{\bf p}+\delta f_{\bf p}$, where $\delta f_{\bf p}$ is a first
order correction with respect to electric field ${\bf E}$.
The Coulomb interaction is considered as a correction to the residual scattering
off impurities. Thus, $\delta f_{\bf p}=\delta f^0_{\bf p}+\delta f^C_{\bf p}$,
where $\delta f^C_{\bf p}$ can be found via a successive approximation to be
applied to the right hand side of Eq.\eqref{q1}, \cite{Yudson}. One can find
\begin{gather}\label{q4}
\delta f^0_{\bf p}=-\tau_e({\bf F}\cdot{\bf v}_{\bf p})
\frac{\partial f_{\bf p}}{\partial \epsilon_{\bf p}},\,\,\,\,
\delta f^C_{\bf p}=\tau_eQ_{eh}\{\delta f^0_{\bf p}\},
\end{gather}
where the electron-hole collision operator being linearized via the first order with
respect to $\delta f^0_{\bf p(k)}$, reads
\begin{gather}\label{q5}
Q_{eh}\{\delta f^{0}_{\bf p}\}=-2\pi\sum_{{\bf p}',{\bf k}',{\bf k}}|U_{{\bf
p}'-{\bf p}}|^2\times\\\nonumber
\Bigl[\delta f^{0}_{\bf p}[(1-n_{\bf k})n_{{\bf k}'}n_{{\bf p}'}+n_{\bf k}(1-
n_{{\bf k}'})(1-n_{{\bf p}'})]\\\nonumber
-\delta f^{0}_{{\bf p}'}[(1-n_{{\bf k}})(1-n_{{\bf p}})n_{{\bf k}'}+n_{\bf
k}n_{\bf p}(1-n_{{\bf k}'})]\\\nonumber
+\delta f^{0}_{\bf k}[(1-n_{\bf p})n_{{\bf k}'}n_{{\bf p}'}+n_{\bf p}(1-n_{{\bf
k}'})(1-n_{{\bf p}'})]\\\nonumber
-\delta f^{0}_{{\bf k}'}[(1-n_{{\bf k}})(1-n_{{\bf p}})n_{{\bf p}'}+n_{\bf
k}n_{\bf p}(1-n_{{\bf p}'})]\Bigr]
\\\nonumber
\times\delta(\epsilon_{{\bf k}'}+\epsilon_{{\bf p}'}-\epsilon_{{\bf k}}-
\epsilon_{{\bf p}})
\delta_{{\bf k}'+{\bf p}'-{\bf k}-{\bf p}}.
\end{gather}
The first order correction to the distribution functions can be presented in the
form ($\varphi_{\bf p}=-\tau_e ({\bf F}\cdot{\bf v}_{\bf p})$)
\begin{gather}\label{q6}
\delta f_{\bf p}=\frac{\partial n_{\bf p}}{\partial \epsilon_{\bf
p}}\varphi_{\bf p}=-\frac{\varphi_{\bf p}}{T}n_{\bf p}(1-n_{\bf p}).
\end{gather}
Taking into account the following auxiliary relations
\begin{gather}\label{q7}
\delta(\epsilon_{{\bf k}'}+\epsilon_{{\bf p}'}-\epsilon_{{\bf k}}-\epsilon_{{\bf
p}})=\\\nonumber
\int d\omega\delta(\epsilon_{{\bf k}'}-\epsilon_{{\bf k}}-
\omega)\delta(\epsilon_{{\bf p}'}-\epsilon_{{\bf p}}+\omega),\,\,\,\,\\\nonumber
\delta_{{\bf k}'+{\bf p}'-{\bf k}-{\bf p}}=\sum_{\bf q}\delta_{{\bf k}'-{\bf k}-
{\bf q}}\delta_{{\bf p}'-{\bf p}+{\bf q}},\\\nonumber
(1-n_{\bf p})n_{{\bf p}'}=N_{-\omega}(n_{\bf p}-n_{{\bf p}'}),\\\nonumber
(1-n_{\bf k})n_{{\bf k}'}=N_{\omega}(n_{\bf k}-n_{{\bf k}'}),\\\nonumber
N_\omega=\frac{1}{e^{\omega/T}-1},\,\,\,\,
\frac{\partial N}{\partial \omega}=\frac{N_{-\omega}N_{\omega}}{T},
\end{gather}
one can find the correction to electron conductivity in the form
%
%
%
%
\begin{gather}\label{q8}
\delta\sigma^e_{xx}=2\pi e^2\frac{\tau_e}{m_{e}}
\left(\frac{\tau_e}{m_e}+\frac{\tau_h}{m_h}\right)\times\\\nonumber
\int\limits_0^\infty\frac{qdq}{2\pi}\frac{q^2|U_{\bf q}|^2}{2}
\int\limits_{-\infty}^{+\infty}
d\omega\frac{
\textmd{Im}\,\Pi^e(-{\bf q},-\omega)\textmd{Im}\,\Pi^h({\bf q},\omega)}
{4\pi^2T\sinh^2(\omega/2T)},
\end{gather}
where
\begin{gather}\label{q9}
\Pi^h({\bf q},\omega)=\sum_{\bf k}
\frac{n_{\bf k}-n_{{\bf k}+{\bf q}}}{\epsilon_{\bf k}-\epsilon_{{\bf k}+{\bf
q}}+\omega+i0}
\end{gather}
is a hole polarization operator. Electron polarization operator $\Pi^e({\bf
q},\omega)$ has a similar form.  Below we consider the quasiballistic and diffusive
hole motion separately.

\textit{Quasiballistic regime of hole motion.---}
In a quasiballistic regime for electrons and holes, the polarization operators
read
\begin{gather}\label{ballistic1}
\textmd{Im}\,\Pi^h({\bf q},\omega)=-
\frac{m_h}{2\pi}\frac{\omega}{v_hq};\,
\textmd{Im}\,\Pi^e({-\bf q},-\omega)=\frac{m_e}{2\pi}\frac{\omega}{v_eq}.
\end{gather}
Substituting these expressions into Eq.\eqref{q8}, taking the integrals over
$\omega$, one finds
%
%
%
%
%
%
\begin{gather}\label{ballistic2}
\delta\sigma^e_{xx}=-\frac{e^2\tau_e T^2}{3(2\pi)^2m_e}
\frac{\tau_e m_h+\tau_h m_e}{v_ev_h}\int\limits_0^{2p_0}qdq|U_{\bf q}|^2,
\end{gather}
where $p_0=\textmd{min}(p_e,p_h)$, and $p_0\gg q_s$. This expression describes
the electron conductivity correction under the quasi-ballistic motion of elections
and holes.

\textit{Diffusive regime of hole motion.---}In the diffusive limit, $D_hq^2\tau_h\ll 1$ and $\omega\tau_h\ll 1$. In this limit
the hole polarization operator is
\begin{gather}\label{diffusive1}
\textmd{Im}\Pi^h({\bf q}, \omega)=-\frac{m_h}{2\pi}\frac{\omega
D_hq^2}{\omega^2+(D_hq^2)^2}.
\end{gather}
Thus, the conductivity correction can be written in the form
\begin{gather}
\delta\sigma^e_{xx}=-
\frac{e^2\tau_e}{(2\pi)^4}\frac{(\tau_em_h+\tau_hm_e)}{p_e}\times\\\nonumber
\times2T\int\limits_0^{1/l_{h}}q^2|U_{\bf q}|^2dq\int\limits_{-
1/2T\tau_h}^{1/2T\tau_h}
\frac{x^2dx}{\sinh^2(x)}\frac{\beta}{(x^2+\beta^2)},
\end{gather}
%
%
%
%
%
%
where $\beta=D_hq^2/2T=(ql_h)^2/2(T\tau_h)$ and $l_h^2=D_h\tau_h$ is a hole
diffusive length. In a diffusive limit $T\tau_h\ll 1$, we have two possible
cases $T\ll D_hq^2\ll 1/\tau_h$ and $D_hq^2\ll T\ll 1/\tau_h$. In the first case
$\beta\gg1$, whereas in the second case $\beta\ll1$. Thus, the integral over
$x$ and the conductivity expression can be estimated for $\beta\gg1$ as
\begin{gather}\label{diffusive3}
\int\limits_{-1/2T\tau_h}^{1/2T\tau_h}
\frac{x^2dx}{\sinh^2(x)}\frac{\beta}{(x^2+\beta^2)}\sim
\frac{1}{\beta}\int\limits_{-\infty}^{\infty}
\frac{x^2dx}{\sinh^2(x)}=\frac{\pi^2}{3\beta},\\\nonumber
\delta\sigma^{e(I)}_{xx}=-\frac{e^2(\tau_eT)(\tau_hT)}{3(2\pi)^2}
\frac{\tau_em_h+\tau_hm_e}{p_el_h^2}\int\limits_0^{1/l_h}dq|U_{\bf q}|^2.
\end{gather}
In the opposite limit, $\beta\ll1$, one finds
\begin{gather}\label{diffusive4}
\int\limits_{-1/2T\tau_h}^{1/2T\tau_h}
\frac{x^2dx}{\sinh^2(x)}\frac{\beta}{(x^2+\beta^2)}\sim
\beta\int\limits_{-\infty}^{\infty}
\frac{dx}{x^2+\beta^2}=\pi,\\\nonumber
\delta\sigma^{e(II)}_{xx}=-\frac{e^2(\tau_eT)}{(2\pi)^3}
\frac{\tau_em_h+\tau_hm_e}{p_e}
\int\limits_0^{1/l_h}q^2dq|U_{\bf q}|^2.
\end{gather}
%
%
%
%
%
%
%
%
%
%
%
%
%
%
%
%
%
%

\textit{Discussion and comparison with the experiment.---}We developed the theoretical description of the temperature-dependent
electron conductivity corrections due to the electron-hole scattering in a 2D
semimetal. The residual resistance at $T=0$ is determined by the electron and
hole scattering via impurities. Thus, the temperature behavior of the sample
resistivity is determined by the temperature dependence of conductivity
corrections found above. It is expressed as
\begin{gather}\label{result1}
\frac{\rho(T)-\rho_0}{\rho_0}=\rho_0\delta\sigma_{xx}(T),
\end{gather}
where $\rho_0=\rho(T=0)$ is the residual sample resistance at zero temperature.

The found conductivity corrections tends to zero at $T=0$. In the diffusive regime of
hole motion, the linear in $T$ behavior Eqs.\eqref{diffusive4} changes into the quadratic-in-$T$ dependence at moderate temperatures, Eq.\eqref{diffusive3}. Estimate
the temperature range, where the linear-in-$T$ conductivity dominates. To do this,
we assume the short-range Coulomb interaction $U_{\bf q}\sim U_0$ and, from
Eqs.\eqref{diffusive3}-\eqref{diffusive4}, we have
\begin{gather}\label{result2}
\frac{\delta\sigma^{e(I)}_{xx}}{\delta\sigma^{e(II)}_{xx}}\sim T\tau_h\ll1.
\end{gather}
Thus, the linear-in-$T$ contribution dominates,
$\delta\sigma^{e(II)}_{xx}\gg\delta\sigma^{e(I)}_{xx}$ in the diffusive limit of
hole motion, $T\tau_h\ll1$.



\begin{figure}[t]
\centering \includegraphics[width=0.85\columnwidth]{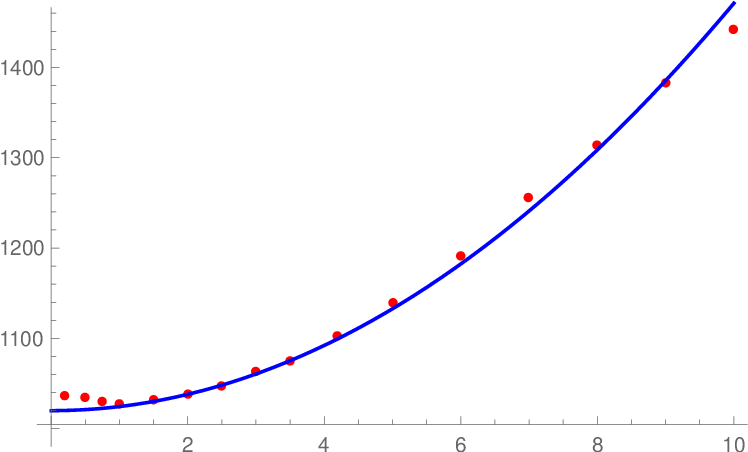}
\caption{Fitting quadratic temperature dependence using Eq.\eqref{exp1}
with parameters $N_e = 4.47*10^{9} cm^{-2}, N_h = 2.3*10^{11}cm^{-2}, \mu_e =
cm^{-1}; \kappa = 25$.}
\label{plot3}
\end{figure}

\begin{figure}[t]
\centering \includegraphics[width=0.85\columnwidth]{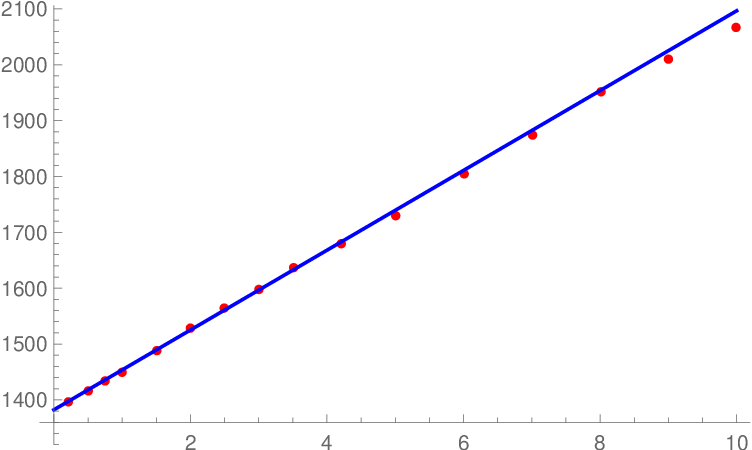}
\caption{Fitting quadratic temperature dependence using Eq.\eqref{exp2}
with parameters $N_e = 7.7*10^{9}cm^{-2}, N_h = 1*10^{10}cm^{-2}, \mu_e
=327535\frac{cm^{2}}{V*s}, \mu_h = 16973\frac{cm^{2}}{V*s}, q_s = 1.14*10^{6}
cm^{-1}, \kappa = 25$.}
\label{plot4}
\end{figure}

Now present a qualitative comparison with the experiment. First of all, we qualitatively account for the quasi-3D geometry of
the real experimental structure.  All the formulas above were obtained in the
assumption that the samples are strictly two-dimensional. Although, in reality, they
are thick ($\sim 20$ nm) and the screening radius could be quite small and even of the
same order as the sample thickness. This may play a
sufficient role in the comparison with the experiment, but the
temperature dependence remains the same $\sim T$ for the diffusive regime and $\sim
T^2$ for the ballistic one. For the ballistic case, one has
\begin{gather}\label{exp1}
\delta\sigma^e_{xx}=-\frac{Ae^2}{3(2\pi)^2\hbar}
\frac{\tau_eT^2(\tau_em_h+\tau_hm_e)}{m_ev_ev_h\hbar^6}
\int\limits_0^{q_1}qdq|U_{\bf q}|^2,
\end{gather}
where $q_1=\textmd{min}(2p_0,1/d)$, and also for the diffusive regime
\begin{gather}\label{exp2}
\delta\sigma^{e(II)}_{xx}=-\frac{Be^2(\tau_eT)}{\hbar(2\pi)^3}
\frac{\tau_em_h+\tau_hm_e}{\hbar^3p_e}
\int\limits_0^{q_2}q^2dq|U_{\bf q}|^2,
\end{gather}
where $q_2=\textmd{min}(1/l_h,1/d)$, $d$ is a QW width, $U_{\bf q}=2\pi
e^2/\kappa(q+q_s)$, and $q_s=\textmd{min}(1/d,1/a_B)$, where $a_B$ is Bohr
radius, and $\kappa$ is a material dielectric constant. Electron and hole
concentrations and mobilities, in both cases, can be easily extracted from the
experimental data. One can find $p_e = \sqrt{2\pi N_e \hbar^2}$ and $p_h=\sqrt{\pi
N_e \hbar^2/2}$, where $N_e$ and $N_h$ are electron and hole concentrations,
respectively. Electron and hole momentum relaxation times can be estimated from
their mobilities as $\tau_e =\mu_e m_e/e$ and $\tau_h = \mu_h m_h/e$,  where
$\mu_e$, $\mu_h$ are electron and hole mobility, respectively. Electron and
hole effective masses are $m_e = 0.025m_0$ and $m_h = 0.15m_0$. Using
Eqs.\eqref{exp1} and \eqref{exp2}, we will fit the experimental data variation of
$A$ and $B$ parameters, Fig.\ref{plot3} and Fig.\ref{plot4}.
Both regimes are fitted via our theoretical expressions Eqs.\eqref{exp1} and
\eqref{exp2} with approximately the same parameter values, $A\approx0.1$ and
$B\approx0.13$. Our theoretical model does not well coincide with the experiment at extremely low temperatures ($T<1K$) in the quadratic-in-T behaviour of semimetal resistivity, Fig.\ref{plot3}. We believe that this discrepancy may be due to the weak-localization scattering of highly mobile electrons off the hole gas subsystem. A theoretical analysis of weak-localization effects is beyond the Boltzmann equation approach developed here. 

\textit{Conclusion.---}We carried out the experimental and theoretical studies of HgTe-based QW structure in the semimetal regime. It was shown that the experimental study demonstrates either linear- or quadratic-in-temperature corrections to the semimetal resistivity depending on the top gate voltage applied to the experimental structure. We suggested a theoretical model and developed the corresponding theoretical description of the experimental system that demonstrates a reasonable agreement with the experiment. 
At the same time, the nonsemimetal domains of gate voltages, close to the charge neutrality point, Fig.\ref{f:4}(a,b), are not described via the presented theoretical model and still require a theoretical grounding.

\textit{Acknowledgements.---}The experimental part of the present study was supported by the Russian Science Foundation, grant No. 23-22-00195. Its theoretical part was financially supported by "BASIS" -- the Foundation for the Advancement of Theoretical Physics and Mathematics.

\end{document}